\documentclass[preprint, pre, showpacs,preprintnumbers,amsmath,amssymb]{revtex4}

\usepackage{graphicx}

\newcommand{\bcD}{\ensuremath{\mathbf{C}}}
\newcommand{\bu}{\ensuremath{\mathbf{u}}}

\newcommand{\ff}{\mbox{\boldmath$f$}}
\newcommand{\rr}{\mbox{\boldmath$r$}}

\newcommand{\uu}{\mbox{\boldmath$u$}}

\newcommand{\bDelta}{\mbox{\boldmath$\Delta$}}

\newcommand{\blambda}{\mbox{\boldmath$\zeta$}}

\begin{document}
\title
{Existence of the $H$ theorem for the athermal lattice Boltzmann models 
with  nonpolynomial equilibria}
\author{Santosh Ansumali}
\email{ansumali@mat.ethz.ch}
\author{Iliya V.\ Karlin}
\affiliation{ETH-Z\"urich, Department of Materials, Institute of Polymers\\
ETH-Zentrum, Sonneggstr. 3, ML J 19, CH-8092 Z\"urich, Switzerland}
\begin{abstract}
The entropy function is found for the 
two-dimensional seven-velocity lattice Boltzmann method on a triangular
lattice. 
Some issues pertinent to the stability and accuracy 
 of the seven velocity lattice Boltzmann method 
are discussed.
\end{abstract}
\pacs{05.20.Dd, 05.70.Ln, 47.11.+j}
\date{\today}
\maketitle

\section{Introduction}

The Lattice Boltzmann method (LBM) is a  useful tool
for  simulations of the complex hydrodynamic phenomena
such as turbulent flows, multiphase flows
and suspensions. It is even believed that the subsequent
development of the lattice Boltzmann method may
provide a new  paradigm in the kinetic modeling due to its mathematical
simplicity and computational efficiency  \cite{SucciBook}.

One important issue, which attracted considerable attention recently, is
the enhancement of stability of the method  \cite{SKC2002}. 
We remind that the early predecessor of the LBM, the lattice gas
method of Frisch, Hasslacher and Pomeau \cite{FHP}, 
was consistent with the $H$ theorem by the microscopic detailed
balance and  was supported by nonpolynomial equilibria
(maximizers of the Fermi-Dirac entropy). 
The method was  was unconditionally stable, but 
in the course of the subsequent development of the LBM
this feature of the unconditional stability, which otherwise would distinguish the 
lattice
Boltzmann method among other methods of computational fluid dynamics,
was gradually lost. The reason why this happened can be traced back to the
earliest versions of the LBM, derived from the lattice gas model, where the true
nonpolynomial equilibria were replaced by their low-order polynomial 
approximations.
Of course, this was motivated by a search for computationally more effective
schemes, beginning with the work of Higuera, Succi and Benzi \cite{Succi89}, 
and which eventually culminated in the athermal single relaxation time 
lattice Bhatnagar-Gross-Krook model  (LBGK) \cite{LBGK1,LBGK2} based on polynomial 
equilibria.
Later studies aimed at restoring the $H$-theorem for the
LBM are well documented (see, e.\ g.\ \cite{SKC2002}).
At present, the entropic LBGK models which
combine computational efficiency of the standard LBGK with the unconditional 
stability pertinent to genuine kinetic models
are constructed \cite{AK02c}.
First results of  simulation of
high Reynolds
number flows confirm the theoretically expected  significant overall gain
in performance of the LBM  by using the  entropic formulations  \cite{KADOS2003}.

The present  study is motivated in part 
by a recent publication \cite{Luo2003} entitled
``Nonexistence of $H$ theorems for the athermal lattice Boltzmann models with
polynomial equilibria''. Therein, the authors demonstrated that 
polynomial equilibria used in the lattice Boltzmann method on the
two-dimensional triangular lattice (the D2Q7 model, see  section \ref{new}) 
do not minimize any convex function, at least when parameters
of these local equilibria are in a certain range.
Since the fact that polynomial equilibria are at odds with 
the maximum entropy principle (and thus they are not local equilibria in
a thermodynamic sense) was pointed out already  for some time (see, for instance,
\cite{Wagner1998}), and actually the experience gained from many studies
of various discrete velocity models \cite{Gatin} does not indicate that polynomials are 
expected as local equilibria, it is therefore 
not surprising that the computation  \cite{Luo2003}  
failed to derive a suitable
entropy function for the D2Q7 model. 

In this paper, we revisit the problem of finding the $H$ function for the
 D2Q7 model. Our approach to derivation of the $H$ function follows the method 
suggested earlier in \cite{K99}.
The straightforward computation presented in detail results in the unique 
Boltzmann-like entropy function for the D2Q7 setup, thereby enabling 
construction of the entropic lattice
Boltzmann models for this lattice.

The structure of the paper is as follows: In section \ref{ELBM},
for the sake of completeness,
we review briefly the athermal LBGK and the athermal entropic LBM, 
in particular, the
entropic LBGK. This section contains no new results. In section \ref{new}, 
we consider the
 2DQ7 lattice, and find the appropriate $H$ function. 
In section \ref{secequilibrium}, 
we find an approximate solution to the equilibrium distribution and
discuss its accuracy. Finally, 
in section \ref{discussion} we put
the entropic lattice Boltzmann method into a perspective with other recent 
approaches
to stabilization of the athermal lattice Boltzmann models.

\section{Lattice Boltzmann and Entropic Lattice Boltzmann}\label{ELBM}
In this section, for the sake of completeness,  
we briefly compare the entropic lattice Boltzmann method (ELBM)
and the standard lattice Boltzmann method (LBM) of athermal hydrodynamics. 

In both the LBM and ELBM methods, one considers populations $f_i$ of discrete
velocities $\mathbf{C}_i$, where $i=1,\dots,m$, at discrete time
$t$. The discrete velocities form the links of a regular and sufficiently 
isotropic lattice,
and it may also include a zero vector.
It is convenient to introduce $m$-dimensional
population vectors $\ff$.  In the isothermal case,
local hydrodynamic variables (density $\rho$  and  momentum density
$\rho\uu$) are defined at lattice sites $\rr$ as:
\begin{align}
\label{Hyd}
\begin{split} \rho = \sum_{i = 1}^{m}\, f_i(\rr, t), \\
\rho\, \uu
= \sum_{i=1}^{m} \mathbf{C}_i\, f_i(\rr, t).
\end{split}
\end{align}
The ELBM begins with finding a convex function of populations, $H$, 
which satisfies the 
following condition: If $f^{\rm eq}_i(\rho,\rho\, \uu)$ minimizes $H$ subject 
to the hydrodynamic
constraints (\ref{Hyd}), then $f^{\rm eq}$ 
also verifies the Galilean invariance of the stress tensor:
\begin{equation}
\label{GI}
  \sum_{i = 1}^{m} C_{i\alpha}\, C_{i\beta}\, f^{\rm eq}_i(\rho,\rho\,
  \uu)=
\rho \, c_{\rm s}^2\delta_{\alpha\beta}+\rho \, u_{\alpha}\, u_{\beta}.
\end{equation}
Here $c_{\rm s}$ is sound speed.

The  $H$ function which satisfies this condition to the 
accuracy of $u^4$, and thus is valid 
to all purposes of incompressible simulations  was derived in \cite{K99} for
the the D1Q3 and the  D2Q9 lattices. Later, this result was extended to the 
three-dimensional D3Q27 lattice
in \cite{AK02a}. Recently, other $H$ functions which verify equation 
(\ref{GI}) to the accuracy
of $u^4$ were found for isotropic Bravais lattices in \cite{Boghosian02}, 
in particular for the
D2Q6 model.
In order to illustrate this, we  list here the results for the $H$ functions 
and their minimizers for the 
${\rm DkQ3}{^{\rm k}}$ lattices
\cite{K99,AK00,AK02,AK02a,AK02b,AK02c}.
Let $D$ be the spatial dimension.  For $D=1$, the three discrete velocities are
$\bcD=\{-1, 0, 1\}$. 
In higher dimensions, the discrete
velocities are tensor products of the discrete velocities of 
these one-dimensional velocities.
The $H$ function is Boltzmann-like:
\begin{equation}
\label{app:H}
H=\sum_{i=1}^{3^D} f_{i}\ln\left(\frac{f_{i}}{w_i} \right).
\end{equation}

Here $w_i$ is the weight associated with the $i$th discrete velocity
$\bcD_i$. For $D=1$, the weights corresponding to the velocities 
$\bcD=\{-1,0,1\}$  are
 $w = \left \{\frac{1}{6},\frac{2}{3}, \frac{1}{6} \right \}$. For $D>1$, 
the weights are constructed by multiplying the weights
associated with each component direction.

The local equilibrium which minimizes (\ref{app:H}) subject to
the fixed  density and momentum reads:
\begin{equation}
\label{TED}
 f^{\rm eq}_i=\rho w_i\prod_{\alpha=1}^{D} 
 \left(2 -\sqrt{1+ 3
 {u_{\alpha}^2}}
 \right)
 \left(\frac{2\,u_{\alpha}
 + 
 \sqrt{1+ 3\,u_{\alpha}^2}}{1-u_\alpha}
 \right)^{C_{i\alpha}}.
\end{equation}
The speed of sound,  $c_{\rm s}$, in this model is $1/\sqrt{\,3}$.
 
Once the entropy function $H$ is found, the basic equation of ELBM to 
be constructed and to be solved is
\begin{equation}
\label{LBM}
f_i(\mathbf{r}+\mathbf{C}_i,t+1)-f_i(\mathbf{r},t)= 
-\beta \alpha[\ff(\mathbf{r},t)]  
\bDelta_i[\ff(\mathbf{r},t)],
\end{equation}
where the right hand side  represents
 the collision process.  
The $m$-dimensional vector function $\bDelta$ (so-called bare 
collision integral),
must satisfy the conditions: 
\[ \sum_{i=1}^m\Delta_i\{1, \mathbf{C}_i\}=0\ 
(\rm{local\ conservation\ laws}),\]
\[ \sigma=\sum_{i=1}^m\Delta_i\frac{\partial H}{\partial f_i}\le 0\ 
(\rm{entropy\ production\ inequality}).\]
Moreover, the local equilibrium vector $\ff^{\rm eq}$ must 
be the only zero point of $\bDelta$, that is,
$\bDelta(\ff^{\rm eq})=\mathbf{0}$, and, finally, $\ff^{\rm eq}$ 
must be the only zero point of the
local entropy production, $\sigma(\ff^{\rm eq})=0$.

The conditions just listed are standard requirements  
taken directly from the well known theory
of the continuous Boltzmann equation. The lattice specifics comes through 
the factor
$\beta \alpha[\ff(\mathbf{r},t)]$ in equation (\ref{LBM}).
Here  $\beta$ is a fixed parameter  in the
interval $[0,1]$,  and is related to the viscosity (the limit $\beta\to 1 $ 
is the zero viscosity limit).  
The  scalar function  $\alpha$   is
the nontrivial root of the  nonlinear equation 
\cite{K98,K99,Boghosian01}:
\begin{equation}
\label{step}
H(\ff)=H(\ff +\alpha\bDelta[\ff]).
\end{equation}
It is the function $\alpha$ which ensures the discrete-time $H$-theorem, 
unlike in the continuous-time
case where it essentially suffices to ensure only the entropy production inequality.
Function $\alpha$  has to be computed numerically on each lattice site at 
each time step. In the fully resolved hydrodynamic limit, when
$\ff\to\ff^{\rm eq}$, the solution $\alpha(\ff)$ tends to its limiting
value $\alpha=2$.

In practice, construction of the bare collision integral $\bDelta$ 
is guided by simplicity.
For the $H$ functions (\ref{app:H}), the local equilibria are given
by explicit formula (\ref{TED}), and thus
the BGK form, $\bDelta=\ff-\ff^{\rm eq}$ becomes available for efficient
numerical realizations.
We further refer to this model as the entropic lattice BGK (ELBGK). 
A gradient single-relaxation time models circumventing the BGK form, 
and which are readily
constructed once just the $H$-function is known, were developed
in \cite{AK00,AK02,AK02a} (see also their discussion in the context of
reaction kinetics, \cite{GK02}).

The standard (nonentropic, second-order polynomial) 
LBM can be considered as a truncation of the ELBM's just discussed.
This truncation is done in three steps. 
First, the local equilibria are replaced by their
second-order in $\uu$ polynomials. For example, for $D=2$, if we expand 
the local equilibrium (\ref{TED}) to second order in $u/c_{\rm s}$, 
we derive the polynomial equilibrium
 of the standard D2Q9 model \cite{LBGK2}. In order to distinguish between the
local equilibria when they are minimizers of appropriate entropy functions 
and the $k$th order
polynomial approximations to them, we denote the latter as 
$\tilde{\ff}^{(k)}$.
Second, instead of nonlinear bare collision integrals
one considers linearized forms, 
$\Delta_i=\sum_{j=1}^{m}A_{ij}(f_j-\tilde{\ff}^{(2)})$. 
The simplest option is the BGK form which becomes always  available, 
$\Delta_i=f_j-\tilde{f}_i^{(2)}$. 
Third, when the latter expression is substituted into the 
right hand side of equation 
(\ref{LBM}), the root of the 
equation (\ref{step}) is replaced by $\alpha=2$. Obviously, these operations 
do not leave the $H$-theorem intact, and it is maybe  not surprising that 
a polynomial cannot be
itself a minimizer of any entropy function anymore, as was demonstrated
for the D2Q7 lattice in \cite{Luo2003}. 
In the subsequent section we find the $H$ function for this lattice without 
assuming a polynomial ansatz for the equilibrium.

\section{$H$ function for the 2DQ7 model}\label{new}

The discrete velocities of the D2Q7 model at each site of a planar 
triangular lattice
consist of  a zero vector $\mathbf{C}_0=\mathbf{0}$, 
and of six vectors of equal length,
$\mathbf{C}_i$, where $i=1,\dots,6$, $\mathbf{C}_i = (
\cos{\left((i-1)\pi/3\right)},\sin{\left((i-1)\pi/3\right)}
)$.  The explicit form of 
the seven-dimensional vectors corresponding
to the $x$ and $y$ components of velocities are as follows:

\begin{align}
\mathbf{C}_x &= \left\{ 0, 1, 1/2,  -1/2, -1, -1/2,   1/2\right\},\\
\mathbf{C}_y &= \frac{\sqrt{3}}{2}\left\{  0, 0, 1, 1, 0, -1, -1\right\} .
\end{align}
Accordingly, the population vector $\ff$  is,
\begin{equation}
\ff=\{f_0,f_1,f_2,f_3,f_4,f_5,f_6\}.
\end{equation}
By symmetry arguments, it is sufficient to seek the $H$ function of the form,
\begin{equation}
\label{Htrial}
H(\ff) = h_0(f_0) + \sum_{i=1}^{6} h(f_i),
\end{equation}
where $h_0(x)$ and $h(x)$ are two  convex functions of one variable to be 
determined.
[We could equally begin with a more general form, $H=\sum_{i=0}^6h_i(f_i)$, 
assuming a separate unknown function for each population. However, the
result would be the same.]

The local equilibrium $\ff^{\rm eq}$ is the minimizer of the function $H$ 
(\ref{Htrial}) subject to the
constraints,
\begin{align}
\label{constraints}
\begin{split}
\rho& = \sum_{i=0}^{6} f^{\rm eq}_i,\\
\rho\, u_x &= \sum_{i=0}^{6} f^{\rm eq}_i C_{i x} =
          f^{\rm eq}_1 - f^{\rm eq}_4 + \frac{1}{2} \left( f^{\rm eq}_2 +
 f^{\rm eq}_6 - f^{\rm eq}_3 - f^{\rm eq}_5 \right),
          \\
\rho \,u_y &= \sum_{i=0}^{6} f^{\rm eq}_i C_{i y} =
        \frac{\sqrt{3}}{2} \left( f^{\rm eq}_2 + f^{\rm eq}_3 - f^{\rm eq}_5 -
 f^{\rm eq}_6 \right).
\end{split}
\end{align}

 Let us denote the inverse of the derivative of $h_0$ and $h$ as
 $\mu_0= \left[  h_0^{\prime} \right]^{-1}$ and $\mu =
 \left[  h^{\prime}\right]^{-1}$. 
Then, the formal solution to the minimization problem reads,
\begin{align}
\label{Lagrange}
\begin{split}
 f^{\rm eq}_0 &= \mu_0 \left( \chi\right ),\\
 f^{\rm eq}_1 &= \mu \left( \chi + \zeta_x\right ),\\
 f^{\rm eq}_2 &=\mu \left( \chi + \frac{1}{2}\zeta_x + \zeta_y \right),
\\
 f^{\rm eq}_3 &= \mu\left( \chi - \frac{1}{2}\zeta_x + \zeta_y \right ),
\\
 f^{\rm eq}_4 &= \mu \left( \chi - \zeta_x\right ),\\
 f^{\rm eq}_5 &= \mu\left( \chi- \frac{1}{2}\zeta_x - \zeta_y \right ),
\\
 f^{\rm eq}_6 &= \mu\left( \chi+ \frac{1}{2}\zeta_x - \zeta_y \right ).
\end{split}
\end{align}
Here $\chi$, $\zeta_x$ and $\zeta_y$ are Lagrange multipliers 
associated with the constraints.

By choosing various  pairs of functions for 
$\mu_0$ and $\mu$   constitutive 
relations for the stress tensors (\ref{GI}) is  satisfied with varying degrees of accuracy,
and the goal is to find such functions $\mu_0$ and $\mu$ for which 
the error reduces to higher orders in the powers of velocity $\uu$.
We rewrite the constitutive relation (\ref{GI}) 
in the form of a discrepancy of the components of the stress tensor as:
\begin{align}
\label{defect}
\begin{split}
T_{xx} &= \rho c_s^2 + \frac{(\rho u_x)^2}{\rho} - \sum_{i=0}^{6} f^{\rm eq}_i
C_{i x}C_{i x},\\
T_{yy} &= \rho c_s^2 + \frac{(\rho u_y)^2}{\rho} - \sum_{i=0}^{6} f^{\rm eq}_i
C_{i y}C_{i y},\\
T_{xy} &= \frac{(\rho u_x) (\rho u_y)}{\rho} - \sum_{i=0}^{6} f^{\rm eq}_i
C_{i x}C_{i y}.
\end{split}
\end{align}
Note that the sound speed is not yet defined in these expressions. In fact, as
we will see  soon, the choice of the sound speed is a solvability condition 
in our procedure.

The next step is crucial \cite{K99}: 
We are going to find such functions $\mu_0$ and $\mu$ which
contain no discrepancy up to the orders $\zeta_x^2$, 
$\zeta_y^2$, and $\zeta_x\zeta_y$. In order to do this, we expand the 
terms 
in equations (\ref{defect}) to relevant orders around the point
$\zeta_x=\zeta_y=0$:
\begin{eqnarray}
\rho(\chi,\blambda)& = &\mu_0(\chi)
+ 6\mu(\chi)
+ \frac{3}{2} \mu''(\chi)\zeta_x^2
+2 \mu''(\chi) \zeta_y^2+O(|\blambda|),\nonumber\\
\rho u_x(\chi,\blambda)&=& 3 \mu'(\chi)\zeta_x+
O(|\blambda|^2),\nonumber\\
\rho u_y(\chi,\blambda)&=&  
2\sqrt{3}\mu'(\chi)\zeta_y+O(|\blambda|),\nonumber\\
\sum_{i=0}^{6}f^{\rm eq}_i(\chi,\blambda)C_{i x}C_{i x}
 &=&
3\mu(\chi)+ \frac{1}{4}\mu''(\chi)\left(\frac{9}{2}\zeta_x^2+2
\zeta_y^2\right)+O(|\blambda|^2),\nonumber\\
\sum_{i=0}^{6} f^{\rm eq}_i(\chi,\blambda)
C_{iy}C_{iy}&=&
 3\mu(\chi) +  \frac{3}{4}\mu''(\chi)
\left(\frac{1}{2}\zeta_x^2 + 2\zeta_y^2\right)+O(|\blambda|^2),\nonumber\\
\sum_{i=0}^{6} f^{\rm eq}_i(\chi,\blambda)
C_{ix}C_{iy}&=&
 \frac{\sqrt{3}}{2}\mu''(\chi)\zeta_x\zeta_y+O(|\blambda|^2).
\label{expansion}
\end{eqnarray}
Here primes denote corresponding derivatives.

Substituting expansions (\ref{expansion})  in equations (\ref{defect}), 
we require that discrepancy (\ref{defect}) vanishes to  second
order in $\blambda$. 

At zero order we have two identical equations:
\begin{eqnarray}
T^{(0)}_{xx}&=&c_{\rm s}^2(\mu_0(\chi)+6\mu(\chi))-3\mu(\chi)=0,
\nonumber\\
T^{(0)}_{yy}&=&c_{\rm s}^2(\mu_0(\chi)+6\mu(\chi))-3\mu(\chi)=0,
\end{eqnarray}
whereupon,
\begin{align}
\label{zeroE}
c_s^2 &=    \frac{3 \mu(\chi)}{\mu_0(\chi) 
+ 6 \mu(\chi)}.
\end{align}
There are no terms linear in $\blambda$ in the expansion of $T_{\alpha
  \,\beta}$.
At second order we get the following equations:
\begin{eqnarray}
T^{(2)}_{xx}&=&
\left\{\left(\frac{3}{2}c_{\rm s}^2-\frac{9}{8}\right)\mu''(\chi)
+\frac{9[\mu'(\chi)]^2}{\mu_0(\chi)+6\mu(\chi)}\right\}
\zeta_x^2+\left(2c_{\rm s}^2-\frac{1}{2}\right)\mu''(\chi)
\zeta_y^2,\label{T2xx}\\
T^{(2)}_{yy}&=&\left(\frac{3}{2}c_{\rm s}^2-\frac{3}{8}\right)\mu''(\chi)
\zeta_x^2+
\left\{\left(2c_{\rm s}^2-\frac{3}{2}\right)\mu''(\chi)
+\frac{12[\mu'(\chi)]^2}{\mu_0(\chi)+6\mu(\chi)}\right\}
\zeta_y^2,\label{T2yy}\\
T^{(2)}_{xy}&=&\frac{\sqrt{3}}{2}\left\{\frac{12[\mu'(\chi)]^2}
{\mu_0(\chi)+6\mu(\chi)}-\mu''(\chi)\right\}\zeta_x\zeta_y.
\label{T2xy}
\end{eqnarray}
We now require $T^{(2)}_{\alpha\,\beta}=0$ independently of the values
of $\blambda$. Thus, setting to zero each term in front of each
combination $\zeta_{\alpha}\zeta_{\beta}$ in equations
(\ref{T2xx}), (\ref{T2yy}), and (\ref{T2xy}), we obtain five equations:
\begin{eqnarray}
\left(\frac{3}{2}c_{\rm s}^2-\frac{9}{8}\right)\mu''(\chi)
+\frac{9[\mu'(\chi)]^2}{\mu_0(\chi)+6\mu(\chi)}=0,\label{xxx}\\
\left(2c_{\rm s}^2-\frac{1}{2}\right)\mu''(\chi)=0,\label{xxy}\\
\left(\frac{3}{2}c_{\rm s}^2-\frac{3}{8}\right)\mu''(\chi)=0,\label{yyx}\\
\left(2c_{\rm s}^2-\frac{3}{2}\right)\mu''(\chi)
+\frac{12[\mu'(\chi)]^2}{\mu_0(\chi)+6\mu(\chi)}=0,\label{yyy}\\
\frac{12[\mu'(\chi)]^2}{\mu_0(\chi)+6\mu(\chi)}-\mu''(\chi)=0.
\label{xy}
\end{eqnarray}
Equations (\ref{xxy}) and (\ref{yyx}) are identical. Assuming
$\mu''\ne0$, equations (\ref{xxy}) and (\ref{yyx}) fix the 
value of sound speed:
\begin{equation}
\label{ss}
c_s=\frac{1}{2}.
\end{equation}
It is straightforward to demonstrate that with the value of sound speed 
(\ref{ss}) the remaining three equations are resolvable.
Indeed, substituting (\ref{ss}) into the zero-order relation (\ref{zeroE}),
we obtain,
\begin{equation}
\label{relation}
\mu_0(\chi)=6\mu(\chi).
\end{equation}
Substituting equations (\ref{ss}) and (\ref{relation})
into equations (\ref{xxx}), (\ref{yyy}), and (\ref{xy}), we find
out that each of the latter three equations reduce to the same
ordinary differential equation:
\begin{equation}
\label{Diff}
 \frac{
[\mu'(\chi)]^2
}{\mu(\chi) } - \mu''(\chi)=0.
\end{equation}
The general solution to this equation is
\begin{equation}
\label{diff:gen}
\mu(\chi) =A \exp{(\chi)}+ B.
\end{equation}
Now,  from the requirement that $\mu''\ne0$ we get  $A\neq0$,
and furthermore $A>0$ by required concavity of $H$.
Substituting the general solution (\ref{diff:gen}) into the 
equation (\ref{Diff}), we find $B=0$. 
 From equation (\ref{relation}) we then have
$\mu_0(\chi) = 6 A\exp{(\chi)}$. Therefore,
$h'(x)=\mu^{-1}(x)=\ln {(x/A)}$, so that $h(x)=x(\ln (x/A)-1)+k_1$, and similarly,
$h_0(x)=x(\ln(x/(6\,A))-1)+k_2$, where $k_1$ and $k_2$ are 
arbitrary constants.
Thus, we obtain the family of Boltzmann-like $H$ function of the 2DQ7 model:
\begin{equation}
\label{result1}
H =f_0\left( \ln\left(\frac{f_0}{6A}\right)-1\right)
 + \sum_{i=1}^{6}  f_i
\left(\ln\left( \frac{f_i}{A}\right) -1\right)+C. 
\end{equation}
As is well known, adding a linear combination of the locally conserved 
quantities to the entropy function is immaterial, so we can fix $A=1/e$, where 
$e$ is the base of natural logarithm,  and $C=0$:
\begin{equation}
\label{result2}
H =f_0 \ln\left(\frac{f_0}{6}\right)
 + \sum_{i=1}^{6}  f_i
\ln\left( f_i\right). 
\end{equation}
In the next section, we shall study the equilibria corresponding to the entropy function (\ref{result2})..

\section{Equilibrium populations}\label{secequilibrium}
  
By construction, the  expansion of the
minimizers of $H$ (\ref{result2}) around the point $\uu=0$ to the order
$u^2$ satisfy all the usual  requirements needed to derive the athermal
Navier-Stokes equations  (cf.\ Ref.\ \cite{K99}, see also below for the present case). 
The exact minimizers are nonpolynomial, and are not always available in a closed form.
Nevertheless, a glimpse of the full
solution is possible since the explicit solution can be computed for a few
special cases. For example, when $u_x = \sqrt{3} \,u_y$, we find  the
exact minimizer of the $H$ function (\ref{result2}),
\begin{align}
\begin{split}
\label{exact:2DQ7}
f_0^{\rm eq} &= \rho \left[ 1 - \frac{1}{6}\sqrt{\left(9 + 48
      u_x^2\right)}\right],\\
f_4^{\rm eq} =f_5^{\rm eq} &=\frac{\rho}{36}
 \left[ 2\sqrt{\left(9 + 48
      u_x^2\right)} -3 -12 u_x\right],\\
f_3^{\rm eq} =f_6^{\rm eq} &=\frac{f^{\rm eq}_0}{6},\\
f_1^{\rm eq} =f_2^{\rm eq} &=\frac{2 \,\rho u_x}{3} + f_4^{\rm eq}.
\end{split}
\end{align}
The exact solution (\ref{exact:2DQ7}) will be used below to test the accuracy of various approximations
to the equilibrium populations.

For practical realizations, we
describe a systematic procedure to obtain the equilibrium  in a
series representation. The procedure relies on the fact that at zero
velocity equilibrium is known exactly,
 \begin{align}
f_i^{\rm eq}(\rho,\mathbf{0}) =\tilde{f}^{(0)}_i= \rho w_i,\ w_0=1/2,\ w_j=1/12, \ j=1,\dots, 6.
\end{align}
Once the exact solution for zero velocity  is known, extension to $\bu\ne0$
is   found by perturbation. 
Specifically, the Lagrange multipliers are expanded as
\begin{align}
\label{expansionL}
\chi &= \sum_{n=0}^{\infty} \epsilon^{n} \chi^{(n)}; \qquad
\zeta = \sum_{n=0}^{\infty} \epsilon^{n} \zeta^{(n)}, \;
\end{align}
where  we have introduced a bookkeeping parameter $\epsilon$, , such that  $u_\alpha
 \to \epsilon \, u_\alpha$,  and $\epsilon$ is set to one in the end of computation. 
This series representation of Lagrange multipliers, when substituted
into equations for the constraints,  induces polynomial approximations $\tilde{f}_i^{(k)}$
of increasingly higher order,
\begin{align}
\tilde{f}_i^{(k)} = \sum_{n=0}^{k} \epsilon^{n}  f_i^{(n)}. 
\end{align}
Further, we seek the expansion parameters consistent with the
conservation constraints at all orders. This translates into a set of
linear equation solved recursively:
\begin{align}
\sum_{i=0}^{m}\tilde{f}_i^{(k)} = \rho, \qquad  \sum_{i=0}^{m} C_{i
  \alpha}\tilde{f}_i^{(k)} = \epsilon \,\rho \, u_{\alpha}, \qquad k\geq 2.
\end{align}
For example, at  first order of this expansion
\begin{align}
\begin{split}
\sum_{i=1}^{6}\, f_i^{(0)}\left[\chi^{(1)} + \zeta_{x}^{(1)} C_{i x} +
  \zeta_{y}^{(1)} C_{i y}\right] &= 0,\\
\sum_{i=1}^{6}\, f_i^{(0)} \, C_{i x} 
\left[\chi^{(1)} + \zeta_{x}^{(1)} C_{i x} + \zeta_{y}^{(1)} C_{i
  y}\right] &= \,\rho\, u_{x}, \,\\
\sum_{i=1}^{6}\, f_i^{(0)} \, C_{i y} 
\left[\chi^{(1)} + \zeta_{x}^{(1)} C_{i x} + \zeta_{y}^{(1)} C_{i
  y}\right] &
= \,\rho\, u_{y}. 
\end{split}
\end{align}
By solving this linear system of three equations we get, 
\begin{align}
\chi^{(1)} &=  0, & \zeta_{x}^{(1)}
&=  4\, u_{x}, & \zeta_{y}^{(1)} &= 4\, u_{y}. 
\end{align}

Similary, at  second order:
\begin{align}
\begin{split}
\sum_{i=1}^{6}\, f_i^{(0)}\left[
 \frac{ 2\, \chi^{(2)} +2 \left( \zeta_{\alpha}^{(2)} C_{i \alpha} \right) +
        {\left( \zeta_{\alpha}^{(1)} C_{i \alpha} \right) }^2
 }{2}\right] &= 0,\\
\sum_{i=1}^{6}\, f_i^{(0)} \, C_{i x} \left[
 \frac{ 2\, \chi^{(2)}+2 \left( \zeta_{\alpha}^{(2)} C_{i \alpha} \right)  + 
        {\left( \zeta_{\alpha}^{(1)} C_{i \alpha} \right) }^2
 }{2}\right] &= 0,\\
\sum_{i=1}^{6}\, f_i^{(0)} \, C_{i y} \left[
 \frac{ 2\, \chi^{(2)} +2 \left( \zeta_{\alpha}^{(2)} C_{i \alpha} \right) + 
        {\left( \zeta_{\alpha}^{(1)} C_{i \alpha} \right) }^2
 }{2}\right] &= 0,
\end{split}
\end{align}

which gives:

\begin{align}
\chi^{(2)} &=- 2 \,( u_x^2 + u_y^2), &  
\zeta_{x}^{(2)} &= 0, &\zeta_{y}^{(2)} &= 0.
\end{align}

Solutions at higher orders 
are easily found using symbolic computation facilities.
In Table \ref{tab:table1}, we present terms of the expansion of the Lagrange multipliers
(\ref{expansionL}) up to  $8$th order in $u$.

 \begin{table}[ht]
\caption{\label{tab:table1}  Expansion coefficients  of 
  Lagrange Multipliers.}
\begin{ruledtabular}
\begin{tabular}{r|c|c|c}
$k$ & $\chi^{(k)}$ &  $\zeta_{x}^{(k)}$ & $\zeta_{y}^{(k)}$\\
\hline 
$1$ & $0$ & $4 \, u_x$ & $4\, u_y$\\
\hline
$2$ & $-2\, \left( u_x^2+ u_y^2\right)$ & $0$ & $0$\\
\hline
$3$ & $0$ & $0$ & $0$\\
\hline
$4$ & $0$ & $0$ & $0$\\
\hline
$5$ & $0$ & $\frac{32}{ 15} u_x^3 \left(  u_x^2 + 5 \,  u_y^2 \right)$
& $ \frac{16}{ 15} u_y \left( 5\, u_x^4 + 3 \,  u_y^4 \right)$\\
\hline
$6$ &$-\frac{8}{9}\, (2 \, u_x^6 + 15 \,u_x^4\,u_y^2 + 3 \,u_y^6)$ & $0$ & $0$\\
\hline
$7$ & $0$ & 
$-\frac{32}{63} \,u_y\,(14\,u_x^6 + 21\,u_x^4\,u_y^2 + 9\,u_y^6)
$
& $-\frac{32}{63}\,(5\,u_x^7 + 42\,u_x^5\,u_y^2 +
21\,u_x^3\,u_y^4)
 $\\
\hline
$8$ &$\frac{4}{9}\,(5\,u_x^8 + 56\,u_x^6\,u_y^2 + 42\,u_x^4\,u_y^4 +
  9\,u_y^8)$ & $0$ & $0$\\
 \end{tabular}
\end{ruledtabular}
\end{table}

This simple procedure  gives the equilibrium distribution at eventually  any
desired order of accuracy. 
For example, the $O(u^6)$ accurate approximation  of the
equilibrium is,
\begin{align}
\begin{split}
\label{feq5}
\tilde{f}_i^{(5)} = \rho w_i \Biggl[&1 +  \zeta_{\alpha}^{(1)} C_{i \alpha} + 
         \frac{ 2\, \chi^{(2)} + 
        {\left( \zeta_{\alpha}^{(1)} C_{i \alpha} \right) }^2 }{2} 
+\frac{ 6\, \chi^{(2)} + 
        {\left( \zeta_{\alpha}^{(1)} C_{i \alpha} \right) }^2 }{6} \,
         \zeta_{\alpha}^{(1)} C_{i \alpha} 
\\
\, &+\frac{ 12\, \chi^{(2)} \left( \chi^{(2)} +  {\left(
         \zeta_{\alpha}^{(1)} C_{i \alpha} \right) }^2\right) +
        {\left( \zeta_{\alpha}^{(1)} C_{i \alpha} \right) }^4 }{24}
\\
\, &+\frac{ 20\, \chi^{(2)}\left(
         \zeta_{\alpha}^{(1)} C_{i \alpha} \right) \left( 3\,\chi^{(2)} +  {\left(
         \zeta_{\alpha}^{(1)} C_{i \alpha} \right) }^2\right) +
        {\left( \zeta_{\alpha}^{(1)} C_{i \alpha} \right) }^5+
\left( \zeta_{\alpha}^{(5)} C_{i \alpha} \right)
 }{120}
\Biggr].
\end{split}
\end{align}

We note in passing that, by construction, each approximate equilibrium population,
$\tilde{\ff}^{(k)}(\rho,\uu)$, satisfies exactly the consistency relations,
$\rho(\tilde{\ff}^{(k)}(\rho,\uu))=\rho$, and $\rho\uu(\tilde{\ff}^{(k)}(\rho,\uu))=\rho\uu$,
at each order $k$.

With the above results, we proceed now to evaluate the higher order
moments of the local equilibrium pertinent to establishing the
hydrodynamic limit of the model.
The components of the equilibrium pressure tensor are:
\begin{align}
\label{pressure2DQ7}
\begin{split}
P_{x \,x}^{\rm eq} &= \rho \left(c_s^2 + {u_x}^2\right)- 
\underbrace{\rho \left[ 
  \,{u_x}^2 \frac{2\, \,
\left({u_x}^2 + 3\, {u_y}^2 \right)}{3} \right] }+ O(u^6),\\
P_{y \,y}^{\rm eq} &= \rho \left(c_s^2 + {u_y}^2\right)-
\underbrace{\rho \left[ 
  \frac{ 
\left({u_x}^4 + 3\, {u_y}^4 \right)}{3} \right] }+ O(u^6),\\
P_{x\, y}^{\rm eq}& =  \rho\,u_x\,u_y - 
\underbrace{\frac{4}{3}\rho\,{u_x}^3 \,u_y}
 + O(u^6),\\
P_{y\, x}^{\rm eq}& =  \rho\,u_y\,u_x- 
\underbrace{\frac{4}{3}\rho\,{u_x}^3 \,u_y}
 + O(u^6).
\end{split}
\end{align}
Here the leading  in $u$  terms  are responsible for  Galilean invariance of the momentum equation,
whereas  the under-braced term give the leading-order deviations. 
All these deviations are of order $O(u^4)$, as expected by construction 
of the $H$ function (\ref{result2}). 
We note in passing that in the D2Q7 case under consideration, 
 the accuracy of the equilibrium stress tensor, as compared with the entropic  2DQ9
model (\ref{TED}), is reduced on two counts: First,  since $c_s^2$ is smaller, 
the  effect of deviations  is larger,  and secondly, 
the off-diagonal part in equation (\ref{pressure2DQ7})
is $O(u^4)$ accurate whereas it is 
 $O(u^6)$ accurate in the entropic 2DQ9 model (\ref{TED}).  
Similarly, the contracted 
third order moment, $Q_{\alpha\beta\beta}^{\rm eq}
=\sum_{i=0}^{6}c_{i\alpha}c^2_if^{\rm eq}_i$, 
related to the equilibrium energy flux, is
\begin{align}
\label{Heat2DQ7}
Q_{\alpha\beta\beta} ^{\rm eq}= (\rho c_{\rm s}^2(D+2)+\rho u^2)u_{\alpha}  -  
\underbrace{\rho u_{\alpha} u^2}.
\end{align}
Again, the first  term in the latter equation correspond to the 
well-known result of the continuous kinetic theory, whereas the under-braced
term is the deviation. Due to the lattice symmetry,
the moment $Q_{\alpha\beta\beta}^{\rm eq}$ is the same for any equilibrium on the D2Q7 lattice,
 and it
is only accurate to
the linear order. As we 
will see it below, it is the accuracy of $Q_{\alpha\beta\beta}^{\rm eq}$
which dictates the working window of the method.

With the expansion method described above, 
one can develop the D2Q7 ELBGK model. 
The bare collision integral (cf.\ section \ref{ELBM})
is assumed in the form, $\Delta_i=-(f_i-\tilde{f}^{(k)}_i)$, where
it should be  decided that uo to which order $k$ is appropriate. 
Three condition guide the choice of the order $k$: 
\begin{itemize}
\item Approximation of the populations should be good enough
to enable solving for the entropy estimate (\ref{step});
\item Deviations in the stress tensor and in the energy flux
should be small.  Since the deviations in the stress tensor 
are one order higher as compared to the energy flow, 
the latter will be most crucial;
\item The velocity window should not be too close to zero
in order to avoid large computational time, and also in order to avoid 
computations with too small numbers.

\end{itemize}

\begin{figure}[ht]
\begin{center}
\includegraphics[scale=0.55]{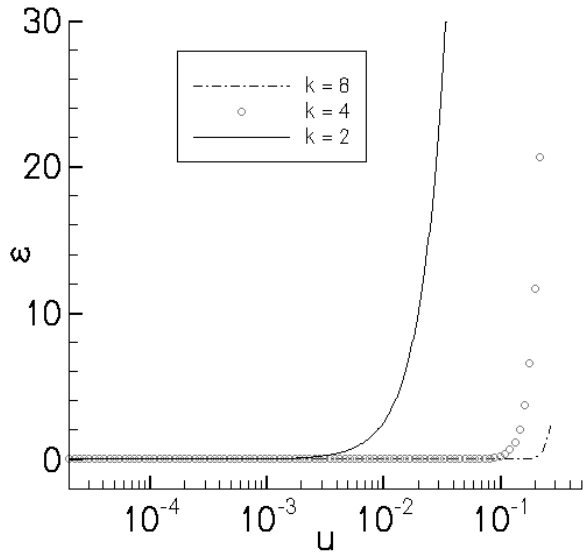}
\caption{\label{Fig1} Deviations of the approximate equilibrium
  population 
  $\tilde{f}_0^{(k)}$ from the exact solution  (see Eq. \eqref{exact:2DQ7} for
  $f_0^{eq}$). Function   $\varepsilon= 10^{5} \times (f_0^{eq} -
  \tilde{f}_0^{(k)})/\rho$ is plotted for three different values of
  $k$.  Notice that for  $k=2$, which correspond to the standard second-order 
polynomial equilibrium used in
  the LBM simulations, the error  starts to increase rapidly 
already  at 
  $u \sim 0.001$. 
All the quantities are given in  dimensionless units.}
\end{center}
\end{figure}

In order to establish the working window and the required order of accuracy  of
the approximation $\tilde{\ff}^{(k)}$ , we test
the results of the expansion against the exact solution (\ref{exact:2DQ7}).
In Fig.\ 
\ref{Fig1}, zero velocity component 
$\tilde{f}^{(k)}_0$ is compared with the exact
solution for various $k$.
It is clear that the usual second-order approximation $\tilde{f}^{(2)}_0$
 is good enough only for $u \leq 0.001$.
However, for a larger velocity window, $u \leq 0.1$, a much better choice
is  to use the $4$th or $6$th order approximation in actual simulations. 
Still higher order approximations do not gain much because
they improve the values of the functions only at velocity too close to the 
sound speed.

\begin{figure}[ht]
\begin{center}
\includegraphics[scale=0.55]{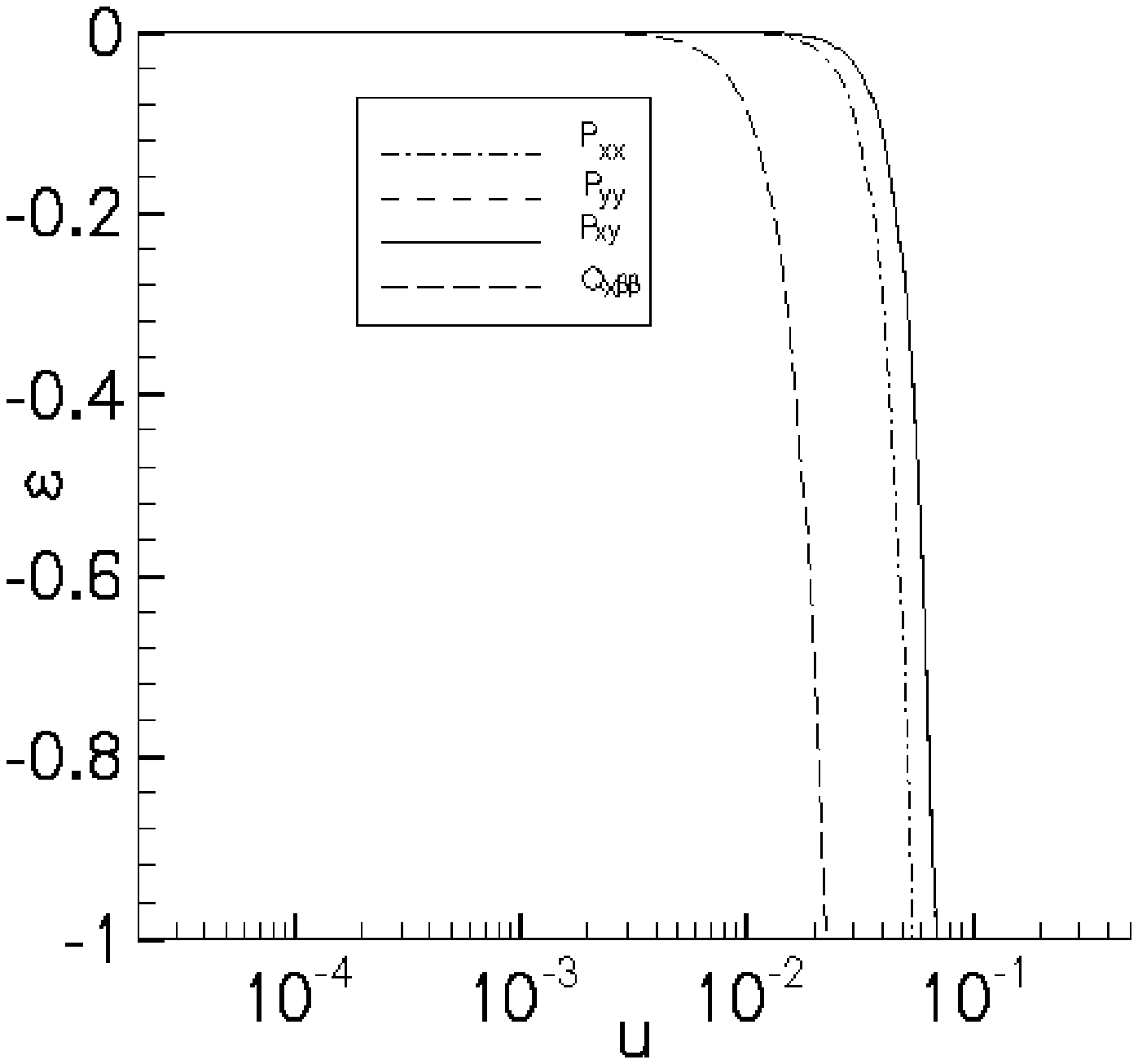}
\caption{\label{Fig2} Deviations  of equilibrium  higher order moments
 $M$. Functions   $\varepsilon=10^{5} \times  ( \Delta M^{\rm
   ELBM})/\rho$ are shown, where, the deviation from the continuous
   case is denoted by $\Delta M^{\rm ELBM}$  and are the same as the
   under-braced term in Eq. \eqref{pressure2DQ7} and Eq. \eqref{Heat2DQ7}.
All the quantities are given in  dimensionless units.}
\end{center}
\end{figure}

Deviations of the stress tensor and of the energy flow 
are demonstrated in 
Fig.\  \ref{Fig2}. This  figure shows that  for
$ u \sim 0.075$, the gain in exactness of the equilibrium populations greatly
outweighs the  minor  deviations  in the pressure tensor. 
On
the other hand,  Fig. \ref{Fig2}  also shows that the dominant deviation 
 is in the energy flow. This error is exactly the same for all 
quadratic approximations employed in the standard LBGK
simulations, and it can be compensated by lowering the 
viscosity.

An alternative to the ELBGK model just described,
is a straightforward realization of the
gradient single relaxation time model
\cite{AK02}, using the $H$ function (\ref{result2}) derived here.
In the present context, construction of the corresponding 
bare collision integral requires only the inversion of a $4\times 4$
matrix with populations-dependent entries which can be done analytically.
This realization does not require any approximation on the equilibrium populations.

\section{Discussion}\label{discussion}
In this paper, we derived the $H$ function for 
the D2Q7 lattice. This makes it possible to derive and implement the
entropic lattice Boltzmann scheme for the triangular lattice, 
in addition to already established models on square lattices.

The goal of the entropic schemes is to achieve nonlinear unconditional
stability in lattice Boltzmann simulations through creation of 
valid kinetic models. We recall that the notion of the  kinetic model
of the Boltzmann equation includes the $H$ theorem as one of the 
important properties \cite{Cercignani,GK94}.
In the construction of entropic lattice Boltzmann models, the
second-order  polynomial approximations
to the generically non-polynomial equilibria of the pertinent $H$ functions  do
appear in the derivations in the same sense as they appeared
in the lattice gas model, that is, to establish theoretically the
hydrodynamic limit to an appreciable degree of accuracy in terms of
the Mach number. 
However, these low-order polynomial approximations do not
show up explicitly in the numerical simulation.
Of course, this analogy should not be misinterpreted,
the entropic lattice Boltzmann equations (\ref{LBM}) are mesoscopic
kinetic equations rather than a lattice gas.
As to the numerical efficiency, for the already existing ELBGK model,
with all the additional burden to solve for the discrete-time
entropy estimate, the serial processor realization requires
only 5 to 10 percent more CPU time as compared to the
usual polynomial LBGK on the 2DQ9 lattice.

Our final comment concerns the so-called multiple relaxation times models
\cite{Lallemand00}.
The idea behind this approach is as follows: 
If one uses a second-order polynomial approximation to the
equilibrium, then not all of the linearized collision integrals of
the form $\Delta_i=\sum_{j=1}^{m}A_{ij}(f_j-\tilde{f}_i^{(2)})$
have the same spectral properties, and one can make use of this to
enhance linear stability by choosing  an appropriate matrix $A_{ij}$.
This is done upon considering spectra of space-dependent problems
(in particular, in a periodic domain \cite{Lallemand00}).
Although the  choice  depends on the boundary
conditions in the specific spectral problem used to determine $A_{ij}$,
it might perform better than the standard LBGK also in other flow situations.

To conclude, it is possible to obtain the $H$ function, and a good
approximation to the correct
equilibria for hydrodynamics on the D2Q7 lattice.  It was shown, that the 
quadratic polynomial  form of the equilibria used in the lattice Boltzmann
method is a good approximation to the correct equilibria  for velocity
$u\sim0.001$.  It is possible to obtain a good approximation to the
equilibria for velocity up to  $u\sim0.1$ by  taking $6$th order approximation to
the correct equilibria.  Further,  lattice Boltzmann simulations
on D2Q7 lattice should avoid using average velocity larger than 
 $u\sim0.075$ due to the dominant errors
in the heat flux. Finally, quadratic polynomials are just a good approximation to
the  equilibria for $u\sim0.001$ and should not be confused with  the
correct equilibria. 

\section*{Acknowledgment}
It is our pleasure to thank Professor A.\ N.\ Gorban,  Professor  H.\ C.\ \"{O}ttinger
 and Professor  S.\ Succi
for useful discussions. Comments of Dr.\ Wen-An Yong are gratefully appreciated.

\end{document}